\documentclass[aps,prx,preprint,letterpaper,groupedaddress,longbibliography]{revtex4-1}
\usepackage[english]{babel}
\usepackage[utf8]{inputenc}
\usepackage{mathtools}
\usepackage{braket}
\usepackage{xcolor}
\usepackage[symbol]{footmisc}
\usepackage{amsmath,amssymb}
\usepackage{caption}
\usepackage{subcaption}
\usepackage{multirow}
\usepackage{scalerel}
\usepackage{tikz}
\usetikzlibrary{svg.path}
\definecolor{orcidlogocol}{HTML}{A6CE39}
\tikzset{
  orcidlogo/.pic={
    \fill[orcidlogocol] svg{M256,128c0,70.7-57.3,128-128,128C57.3,256,0,198.7,0,128C0,57.3,57.3,0,128,0C198.7,0,256,57.3,256,128z};
    \fill[white] svg{M86.3,186.2H70.9V79.1h15.4v48.4V186.2z}
                 svg{M108.9,79.1h41.6c39.6,0,57,28.3,57,53.6c0,27.5-21.5,53.6-56.8,53.6h-41.8V79.1z M124.3,172.4h24.5c34.9,0,42.9-26.5,42.9-39.7c0-21.5-13.7-39.7-43.7-39.7h-23.7V172.4z}
                 svg{M88.7,56.8c0,5.5-4.5,10.1-10.1,10.1c-5.6,0-10.1-4.6-10.1-10.1c0-5.6,4.5-10.1,10.1-10.1C84.2,46.7,88.7,51.3,88.7,56.8z};}}
\newcommand\orcidicon[1]{\href{https://orcid.org/#1}{\mbox{\scalerel*{
\begin{tikzpicture}[yscale=-1,transform shape]
\pic{orcidlogo};
\end{tikzpicture}
}{|}}}}
\usepackage[colorlinks,citecolor=red,urlcolor=blue,bookmarks=false,hypertexnames=true]{hyperref} 

\newcommand{\kwv}{\boldsymbol{k}}
\newcommand{\qwv}{\boldsymbol{q}}
\newcommand{\pwv}{\boldsymbol{p}}
\newcommand{\knorm}{k}
\newcommand{\qnorm}{k}
\newcommand{\pnorm}{p}
\newcommand{\km}{k_m}
\newcommand{\kmnorm}{k_m}
\newcommand{\kangle}{\theta_{\kwv, \kwv^\prime}}

\newcommand{\fkm}{f_{m\kwv}}

\newcommand{\eunits}{kVcm$^{-1}$}
\newcommand{\perturbo}{\textsc{Perturbo}}

\begin{document}
\title{Transport and noise of hot electrons in GaAs using a  semi-analytical model of two-phonon polar optical phonon scattering}

\author{Jiace Sun
\orcidicon{0000-0002-0566-2084}}
\author{Austin J. Minnich \orcidicon{0000-0002-9671-9540}}
\email{aminnich@caltech.edu}

\affiliation{Division of Engineering and Applied Science, California Institute of Technology, Pasadena, CA, USA}

\date{\today}

\begin{abstract}
Recent ab-initio studies of electron transport in GaAs have reported that electron-phonon (e-ph) interactions beyond the lowest order play a fundamental role in charge transport and noise phenomena.
Inclusion of the next-leading-order process in which an electron scatters with two phonons was found to yield good agreement for the high-field drift velocity, but the characteristic non-monotonic trend of the power spectral density of current fluctuations (PSD) with electric field was not predicted.
The high computational cost of the ab-initio approach necessitated various approximations to the two-phonon scattering term, which were suggested as possible origins of the discrepancy.
Here, we report a semi-analytical transport model of two-phonon electron scattering via the Fröhlich mechanism, allowing a number of the approximations in the ab-initio treatment to be lifted while retaining the accuracy to within a few percent.
We compare the calculated and experimental transport and noise properties as well as scattering rates measured by photoluminescence experiments. 
We find quantitative agreement within $15$\% for the drift velocity and $25\%$ for the $\Gamma$ valley scattering rates, and agreement with the $\Gamma-L$ intervalley scattering rates within a factor of two. Considering these results and prior studies of current noise in GaAs, we conclude that the most probable origin of the non-monotonic PSD trend versus electric field is the formation of space charge domains rather than intervalley scattering as has been assumed.

\end{abstract}

\maketitle

\section{Introduction}

Electron transport in semiconductors is of fundamental interest and of high relevance for microelectronic devices \cite{sze2007, Lundstrom_2000, Hartnagel_2001}. The upper limit for the mobility of a semiconductor is governed by scattering of electrons by phonons. Early studies of charge transport properties employed a semi-empirical description of the band structure and electron-phonon scattering. The introduction of the Monte Carlo method allowed for the numerical simulation of transport with fewer approximations \cite{reggiani_1983}. Later, full-band MC tools capable of simulating realistic device geometries were developed, but the treatment of the e-ph scattering rates in general required some fitting parameters.  \cite{ Hess:1991,Fischetti:1991, fischetti_1991_2}.  The development of the ab-initio description of the electron-phonon interactions based on density functional theory (DFT), density functional perturbation theory (DFPT) and Wannier interpolation has enabled the parameter-free computation of low-field charge transport properties such as mobility  \cite{Bernardi_2016, Giustino_2017}. These methods have now been applied to a range of semiconductors, including Si \cite{restrepo2009, Ponce_2018}, GaN \cite{ponce2019gan}, GaAs \cite{Zhou_2016}, two-dimensional materials \cite{borysenko2010, kaasbjerg2012, 2dmobi2020}, and others. Recent methodology developments, including two-phonon scattering \cite{lee2020}, quadrupole interactions \cite{brunin2020,quadprb2020}, and $GW$ corrections \cite{Ponce_2018,gwpt_2019}, have facilitated a rigorous comparison of the accepted level of theory with experiment.

The accuracy of the ab-initio calculations has been mainly based on comparison to low-field mobility values. Recent works have extended these calculations to beyond low-field transport and noise properties. \cite{hatanpaa2022two,warmelectrons,cheng2022high,catherall2023high}
In GaAs, it was found that although the qualitative shape of the drift velocity versus electric field curve was predicted correctly compared to experiment, the magnitude of the drift velocity was overpredicted by about 50\%.\cite{cheng2022high} The inclusion of the next-leading-order term of scattering involving two phonons (2ph)  yielded a low-field mobility and drift velocity with substantially improved agreement. However, the characteristic nonmonotonic trend of PSD with electric field was not predicted even with the 2ph theory. Owing to the high cost of the ab-initio calculations, the treatment of 2ph processes in that work required several approximations, such as the neglect of off-shell 2ph scattering processes.  Whether these neglected processes or other numerical considerations can account for the PSD discrepancy remains unknown.

Here, we introduce a semi-analytical model for both 1ph and 2ph e-ph scattering via the Fröhlich mechanism, allowing the full 2ph scattering term to be treated over the wide range of energies needed for high-field transport while introducing error on the order of only a few percent. We find that the transport and noise properties are qualitatively unchanged compared to the ab-initio calculations. The calculated scattering rates agree with those obtained from photoluminescence experiments to  within 25\% for the $\Gamma$ valley rates, and a factor of two for the $\Gamma$-L intervalley rates. Despite this degree of agreement, the qualitative discrepancy observed previously for the PSD remains. We consider the remaining approximations in the semi-analytical model and find that they are unlikely to account for the PSD discrepancy. Therefore, we conclude that the characteristic peak in the PSD with electric field arises from the formation of space charge domains rather than intervalley scattering as has been assumed in the literature. This finding has implications for the use of transport measurements to study intervalley scattering.

\section{Theory}

\subsection{Overview of formalism for charge transport and noise properties}

We first review the ab-initio treatment of electron transport and electronic noise using the BTE \cite{hamaguchi2010basic}.
For a spatially homogeneous system with electric field $\vec{\mathcal{E}}$, the electron distribution function $\fkm$
is governed by

\begin{equation}\label{eq:BTE}
    \frac{\partial f_{m\kwv}}{\partial t} + \frac{e\vec{\mathcal{E}}}{\hbar}\cdot \nabla_{\kwv}\fkm = \mathcal{I}[\fkm],
\end{equation}

where $f_{m\kwv}$ is the electron occupation in the state with band index $m$ and wave vector $\kwv$, $e$ is the fundamental charge, and $\mathcal{I}[\fkm]$ is the collision term, which describes the scattering of electrons with phonons \cite{Bernardi_2016}.
In this work, index $m$ is dropped in all the following derivations for simplicity as only one band is relevant for electron transport in GaAs in the range of electric fields considered.

For non-degenerate electrons, the collision term can be linearized as \cite{cheng2022high}:

\begin{equation} \label{eq:linear_BTE}
\mathcal{I}[f_{\kwv^\prime}] \approx  - \sum_{\kwv}  \Theta_{\kwv^\prime, \kwv} f_{\kwv},
\end{equation}
where $\Theta_{\kwv^\prime, \kwv}$ is e-ph collision matrix. The diagonal elements $\Theta_{\kwv, \kwv}$ are equal to the total scattering rates as $\Theta_{\kwv, \kwv} = \Gamma_{\kwv} = - \sum_{\kwv^\prime\neq \kwv} \Theta_{\kwv^\prime, \kwv}$. With this linearization and a finite difference representation of the  derivative operator $\nabla_{\kwv}$ \cite{warmelectrons}, Eq.~\ref{eq:BTE} becomes a linear partial-differential equation which can be solved by numerical linear algebra. The equation for the steady distribution function  $f_{\kwv}^s$ is given by Eq.~6 of Ref. \onlinecite{warmelectrons}. Steady-state mean transport properties such as drift velocity can be calculated  with the appropriate Brillouin zone sum using this distribution.

The current power spectral density (PSD) is used to characterize fluctuations in occupation about the mean distribution. The PSD is defined as the Fourier transform of the autocorrelation of the current density fluctuations (Eq. 12 of Ref.~\onlinecite{warmelectrons}).
Following Ref.~\onlinecite{warmelectrons}, the current PSD at frequency $\omega$ can be calculated as
\begin{equation}
\label{noisesum}
    S_{j_{\alpha}j_{\beta}}(\omega) =
    2 \bigg(\frac{2 e}{\mathcal{V}_0}\bigg)^2 \Re \left[ \sum_{\mathbf{k}} v_{\kwv\alpha} g_{\kwv\beta} \right],
\end{equation}
where $v_{\kwv\alpha}$ is the group velocity of the electron at state $\kwv$ along axis $\alpha$, $\mathcal{V}_0$ is the supercell volume, and $g_{\kwv\beta}$ is the effective distribution function \cite{rustagi2014} which satisfies the following equation:

\begin{equation} \label{eq:linear_g}
    \sum_{\kwv} (\Theta_{\kwv^\prime, \kwv} + \frac{e \vec{\mathcal{E}}}{\hbar} \cdot \vec{D}_{\kwv^\prime, \kwv} + i\omega \delta_{\kwv^\prime, \kwv}) g_{\kwv \alpha} = f_{\kwv^\prime}^s (v_{\kwv^\prime,\alpha} - V_{\alpha}),
\end{equation}
where $D$ is the finite difference representation of $\nabla_{\kwv}$ specified in Eq.~\ref{eq:BTE}, $f_{\kwv}^s$ is the steady-state occupation for the state at wave vector $\kwv$, and $V_\alpha$ is the drift velocity along axis $\alpha$.

The e-ph collision matrix is obtained from perturbation theory in orders of the e-ph interaction strength. For 1ph scattering of non-degenerate electrons,  the non-diagonal scattering matrix elements are given by \cite{cheng2022high}:
\begin{equation} \label{eq:Theta_1ph}
    \Theta_{\kwv^\prime = \kwv+\qwv, \kwv}^{\mathrm{(1ph)}} = -\frac{2 \pi}{\hbar} \frac{1}{N} \sum_{\nu}  \left|g_{\nu}(\kwv,\qwv)\right|^2 
     \bigg(\delta(\epsilon_{\kwv}-\hbar \omega_{\nu \qwv}-\epsilon_{\kwv+\qwv})N_{\qwv} + \delta(\epsilon_{\kwv} + \hbar \omega_{\nu\qwv} -\epsilon_{\kwv+\qwv}) (N_{\qwv}+1) \bigg)
\end{equation}
for $\kwv \neq \kwv^\prime$, where $g_{\nu}(\kwv,\qwv)$ is the e-ph scattering matrix element, $\epsilon_{\kwv}$ is the energy of the electronic state $\kwv$, $\omega_{\nu \qwv}$ is the frequency of phonon with mode $\nu$ and wave vector $\qwv$, and $N$ is the total number of $\kwv$ points in the Brillouin zone, $N_{\qwv}$ is the phonon occupation according to the Bose-Einstein statistics. The two delta functions are energy conservation conditions for the emission and absorption subprocesses, respectively.

\subsection{Two-phonon scattering}

The collision integral for 2ph scattering was derived in Ref.~\onlinecite{lee2020}, and the linearized form is given in Ref.~\onlinecite{cheng2022high}. 2ph scattering can be divided into the two-emission (2e), one-emission-one-absorption (1e1a) and two-absorption (2a) subprocesses. Here, we rewrite the original formalism in Ref.~\onlinecite{lee2020} to facilitate the derivations in the next section.
Assuming non-degenerate electron statistics in Eq.~4 in Ref.~\onlinecite{lee2020}, we rewrite Eq.~12 in Ref.~\onlinecite{cheng2022high} by splitting $W^{(i)}$ to two parts:
\begin{equation}
\begin{aligned} \label{eq:old_W}
    W^{(i)} &= |\tilde{W}_{\kwv, \qwv, \pwv, \alpha_2} + \tilde{W}_{\kwv, \pwv, \qwv, \alpha_1}|^2
    \\
    &= \left[|\tilde{W}_{\kwv, \qwv, \pwv, \alpha_2}|^2 + \text{Re}(\tilde{W}_{\kwv, \qwv, \pwv, \alpha_2} \tilde{W}_{\kwv, \pwv, \qwv, \alpha_1}^*)\right] + \left[ |\tilde{W}_{\kwv, \pwv, \qwv, \alpha_1}|^2 + \text{Re}(\tilde{W}_{\kwv, \pwv, \qwv, \alpha_1}\tilde{W}_{\kwv, \qwv, \pwv, \alpha_2}^* ) \right],
\end{aligned}
\end{equation}
where
\begin{equation} \label{eq:W_tilde}
    \tilde{W}_{\kwv, \qwv, \pwv, \alpha} = \frac{ g_{\nu}(\kwv,\qwv)g_{\mu}(\kwv+\qwv,\pwv) }
    { \epsilon_{\kwv'}-\epsilon_{\kwv+\qwv}+\alpha \hbar \omega_{\nu\pwv} + i\eta - i\hbar \Gamma_{\kwv+\qwv} / 2 },
\end{equation}
where $\eta$ is a positive infinitesimal and $\Gamma_{\kwv + \qwv} = \Gamma^{(\mathrm{1ph})}_{\kwv + \qwv} + \Gamma^{(\mathrm{2ph})}_{\kwv + \qwv}$ is the total scattering rate of the intermediate state $\kwv + \qwv$.
For the 2e and 2a subprocesses, the two terms give the same contribution after the summation.
For the 1e1a subprocess, the two terms in Eq.~\ref{eq:old_W} physically represent the emission-then-absorption (a-e) and absorption-then-emission (e-a) subprocesses, respectively.
Finally, we exchange the summation order of $\qwv$ and $\pwv$ in Eq.~9 in Ref.~\onlinecite{cheng2022high} for the second term of  Eq.~\ref{eq:old_W} and arrange the equations to obtain:

\begin{equation} \label{eq:Theta_2ph}
    \Theta^{(\mathrm{2ph})}_{\kwv^\prime, \kwv} = - \frac{2\pi}{\hbar}\frac{1}{N^2} \sum_{\alpha_1 = \pm 1} \sum_{\alpha_2 = \pm 1} \sum_{\qwv + \pwv = \kwv' -\kwv}\sum_{\nu\mu} \tilde{\Theta}^{(\alpha_1, \alpha_2)}_{\kwv,\qwv\nu,\pwv\mu}
\end{equation}

for $\kwv \neq \kwv^\prime$, where $\alpha_1$ and $\alpha_2$ indicate whether the first and second phonon is emitted ($\alpha_{1,2}=1$) or absorbed ($\alpha_{1,2}=-1$), so that the four combinations of $\alpha_{1,2}=\pm 1$ describe the four subprocesses.

The term $\tilde{\Theta}^{(\alpha_1, \alpha_2)}_{\kwv,\qwv\nu,\pwv\mu}$ in Eq.~\ref{eq:Theta_2ph} is defined as 

\begin{equation} \label{eq:Theta_2ph_subprocess}
    \tilde{\Theta}^{(\alpha_1, \alpha_2)}_{\kwv,\qwv\nu,\pwv\mu} = (N_{\qwv} + \delta_{\alpha_1, 1}) (N_{\pwv} + \delta_{\alpha_2, 1})  \left[|\tilde{W}_{\kwv, \qwv, \pwv, \alpha_2}|^2 + \text{Re}(\tilde{W}_{\kwv, \qwv, \pwv, \alpha_2} \tilde{W}_{\kwv, \pwv, \qwv, \alpha_1}^*)\right] \delta 
    (\epsilon_{\kwv} - \epsilon_{\kwv'} - \alpha_{1}\omega_{\nu\qwv} - \alpha_{2}\omega_{\mu\pwv}).
\end{equation}

As the 2ph scattering rates depend on the intermediate state rates, the 2ph scattering must be calculated iteratively.
In each iteration, the intermediate state rates of the previous iteration is used to update the 2ph scattering matrix $\Theta^{(\mathrm{2ph})}_{\kwv^\prime, \kwv}$ and scattering rates $\Gamma^{(\mathrm{2ph})}_{\kwv}$.

Compared with 1ph scattering, 2ph scattering is much more computationally expensive, particularly for high-field transport which requires a larger energy window than for low-field transport. In Ref.~\onlinecite{cheng2022high}, several approximations were made to make the computation feasible, including limiting the number of self-consistent iterations of the 2ph rates to three, restricting the off-shell extent $|\epsilon_{\kwv'}-\epsilon_{\kwv+\qwv}+\alpha \hbar \omega_{\nu\pwv}|$ to 25 meV, and neglecting the interference term $\text{Re}(\tilde{W}_{\kwv, \qwv, \pwv, \alpha_2} \tilde{W}_{\kwv, \pwv, \qwv, \alpha_1}^*)$ term in Eq.~\ref{eq:Theta_2ph_subprocess}.  Additionally, the maximum grid density that could be used was $200 \times 200 \times 200$. The effect of these approximations on the observable transport and noise properties was not assessed. In particular, the on-shell approximation neglects off-shell processes and thus underestimates the scattering rates. These approximations were mentioned as possible reasons for the PSD discrepancy in Ref. \cite{cheng2022high}.

\subsection{Semi-analytical model for 1ph and 2ph $\mathbf{\Gamma}$-$\mathbf{\Gamma}$ scattering} \label{sec:semi-analytical_model}

In this section, we introduce a semi-analytical model to treat 1ph and 2ph $\Gamma-\Gamma$ intravalley scattering by the Fröhlich interaction that retains the accuracy of the ab-initio formalism to within a few percent while allowing the approximations described above to be lifted. 
This model is based on the fact that over the range of wavevectors considered in this study, the $\Gamma$ valley in GaAs is nearly spherically symmetric, and $\Gamma\text{-}\Gamma$ scattering can be accurately described by using only the Fröhlich interaction \cite{verdi2015frohlich}.
The model is valid only for $\Gamma$ intravalley scattering because $\Gamma$-L intervalley scattering lacks an analytic description of similar accuracy.

The semi-analytical model uses the following approximations. First, the band structure is described using the Kane model \cite{kane1957band} for a spherically symmetric, non-parabolic band. This description is accurately satisfied for the $\Gamma$ valley, with the Kane model bands deviating from the ab-initio band structure by at most 7\% over the range of wave vectors considered ($\sim 0.1 G$, where  $G$ is the reciprocal lattice constant). Second, prior works have shown that $\Gamma\text{-}\Gamma$ e-ph scattering in GaAs is dominated by longitudinal optical (LO) phonons via the Fröhlich interaction \cite{frohlich1954electrons}. We therefore neglect  scattering processes involving other phonon branches and scattering mechanisms. The computed matrix elements $g_\text{LO}(\kwv, \qwv)$  for Fröhlich scattering are found to exhibit negligible anisotropy so that $g_\text{LO}(\kwv, \qwv) = g_\text{LO}(\qnorm)$, enabling an analytic form of $g_\text{LO}(\qnorm)$ to be fit to the ab initio values as described  in Sec.\ref{sec:details}. In the range of wave vectors considered, this approximation is satisfied to within 3\% \cite{verdi2015frohlich}. Third, we take the LO phonon frequency to be a constant $\omega_\text{LO} = 35$ meV. In the range of phonon wave vectors $\qnorm \in (0, 0.2G)$ considered here, this assumption is satisfied to within less than 0.3\%.


We now discuss the treatment of 1ph and 2ph e-ph scattering based on these simplifications. The summation in Eq.~\ref{eq:Theta_2ph} may be rewritten as an integral in the Brillouin zone over the intermediate wave vector $\km$ by letting $\qwv \rightarrow \km - \kwv$ and $\pwv \rightarrow \kwv^\prime - \km$.
Additionally, we exploit spherical symmetry to rewrite all the quantities in spherical coordinates as: $\Theta_{\kwv^\prime, \kwv} = \Theta(\knorm, \knorm^\prime, \kangle)$ and $\Gamma_{\kwv} = \Gamma_{\knorm}$. After some simplifications, we obtain the 1ph and 2ph collision matrices as:

\begin{equation} \label{eq:Theta_1ph_integral}
    \Theta^{\mathrm{(1ph)}}(\knorm, \knorm^\prime, \kangle) = \frac{2 \pi}{\hbar} \frac{1}{\Omega_\text{BZ}}  |g_{\text{LO}}(|\kwv^\prime-\kwv|)|^2 
    \sum_{\alpha=\pm1} A_{\alpha}
     \delta(\epsilon_{\knorm}-\alpha \hbar \omega_\text{LO}-\epsilon_{\knorm^\prime})
\end{equation}
and
\begin{equation} \label{eq:Theta_2ph_integral}
    \Theta^{(\mathrm{2ph})}(\knorm, \knorm^\prime, \kangle) = \frac{2\pi}{\hbar}\frac{1}{\Omega_{\text{BZ}}^2} \sum_{\alpha_1 = \pm 1} \sum_{\alpha_2 = \pm 1}  
    A_{\alpha_1}A_{\alpha_2} \delta  (\epsilon_\knorm - \epsilon_{\knorm^\prime} - (\alpha_{1} + \alpha_{2}) \hbar \omega_\text{LO}) I^{(\alpha_1, \alpha_2)}(\knorm, \knorm^\prime, \kangle),
\end{equation}
where $\Omega_{\text{BZ}}$ is the Brillouin zone volume. $A_{\alpha}$ is the phonon occupation factor defined as

\begin{equation} \label{eq:phonon_occupation}
    A_{\alpha}=N_{\text{LO}} + \delta_{\alpha, +1},
\end{equation}
 where $N_{\text{LO}}=(\exp(\hbar \omega_{\text{LO}}/k_B T)-1)^{-1}$ is the LO phonon occupation, and $I^{(\alpha_1, \alpha_2)} = I_\text{1}^{(\alpha_1, \alpha_2)} + I_\text{2}^{(\alpha_1, \alpha_2)}$ is decomposed to the non-interference part $I_\text{1}^{(\alpha_1, \alpha_2)}$ and the interference part $I_\text{2}^{(\alpha_1, \alpha_2)}$:

\begin{equation} \label{eq:int_I}
\begin{aligned}
    I_\text{1}^{(\alpha_1, \alpha_2)}(\knorm, \knorm^\prime, \kangle) &= 
    \int |\tilde{W}_{\kwv, \km - \kwv, \kwv^\prime-\km, \alpha_2}|^2   \mathrm{d}^3 \km
    \\
    &= \int \left|\frac{ g_\text{LO}(|\km-\kwv|)g_\text{LO}(|\kwv^\prime-\km|) }{ \epsilon_{\knorm^\prime}-\epsilon_{\kmnorm}+\alpha_2 \omega_\text{LO} + i\eta - \hbar \Gamma_{\kmnorm} / 2 }\right|^2 \mathrm{d}^3 \km,
\end{aligned}
\end{equation}
and
\begin{equation} \label{eq:int_I_itf}
\begin{aligned}
    I_\text{2}^{(\alpha_1, \alpha_2)}(\knorm, \knorm^\prime, \kangle) &= 
    \int_{\qwv + \pwv = \kwv^\prime - \kwv} \text{Re}(\tilde{W}_{\kwv, \qwv, \pwv, \alpha_2} \tilde{W}_{\kwv, \pwv, \qwv, \alpha_2}^{*})   \mathrm{d}^3 \pwv \mathrm{d}^3 \qwv
\end{aligned}
\end{equation}

Equation \ref{eq:int_I} can be further simplified by writing the integration in spherical coordinates and separating the radius and angular part:

\begin{equation} \label{eq:int_I_spherical}
    I_\text{1}^{(\alpha_1, \alpha_2)}(\knorm, \knorm^\prime, \kangle) = \int \frac{ \tilde{I}^{(\alpha_1, \alpha_2)}(\knorm, \knorm^\prime, \kangle, \kmnorm) }{ |\epsilon_{\knorm^\prime}-\epsilon_{\kmnorm}+\alpha_2 \omega_\text{LO} + i\eta - \hbar \Gamma_{\kmnorm} / 2|^2 } \kmnorm^2 \mathrm{d} \kmnorm ,
\end{equation}
where $\tilde{I}_\text{1}^{(\alpha_1, \alpha_2)}$ is the angular part defined as:
\begin{equation} \label{eq:I_tilde}
     \tilde{I}_\text{1}^{(\alpha_1, \alpha_2)}(\knorm, \knorm^\prime, \kangle, \kmnorm) = \int  |g_\text{LO}(|\km-\kwv|)g_\text{LO}(|\kwv^\prime-\km|)|^2 \sin \theta_{\km} \mathrm{d} \theta_{\km} \mathrm{d} \phi_{\km},
\end{equation}
where $\theta_{\km}$, $\phi_{\km}$ are the polar angle and azimuthal angle defining the intermediate wave vector $\km$, respectively.
Since $\tilde{I}_\text{1}^{(\alpha_1, \alpha_2)}$ is independent of the band structure and the self-energy, recomputation of this term in each 2ph iteration is not required.
In practice, to significantly reduce the computational cost, $\tilde{I}_\text{1}^{(\alpha_1, \alpha_2)}(\knorm, \knorm^\prime, \kangle, \kmnorm)$ are precomputed on a grid of $\knorm$, $\knorm^\prime$, $\kangle$ and $\kmnorm$ before the 2ph iteration. Here $\knorm$, $\knorm^\prime$, $\kangle$ are not independent of  each other due to the energy conservation condition.
Once $\tilde{I}_\text{1}^{(\alpha_1, \alpha_2)}(\knorm, \knorm^\prime, \kangle, \kmnorm)$ is computed on a predefined grid, $\Theta^{(\mathrm{2ph})}(\knorm, \knorm^\prime, \kangle)$ can be calculated according to Eqs.~\ref{eq:Theta_2ph_integral}, \ref{eq:int_I_spherical} and \ref{eq:int_I_itf}.
We note that such separation of the radius and spherical part is not valid for $\tilde{I}_\text{2}^{(\alpha_1, \alpha_2)}$, so an expensive iterative update is required.
However, since $\tilde{I}_\text{2}^{(\alpha_1, \alpha_2)}$ is generally much smaller compared with $\tilde{I}_\text{2}^{(\alpha_1, \alpha_2)}$, we update $\tilde{I}_\text{2}^{(\alpha_1, \alpha_2)}$ every 10 iterations to decrease the cost of the self-consistent calculations.

To complete the 2ph iteration, the last quantities to be computed are the total 1ph and 2ph scattering rates:
\begin{equation} \label{eq:rates_int}
\Gamma_{\knorm}^{(\text{type})} = \int \Theta^{(\text{type})}(\knorm, \knorm^\prime, \kangle) \mathrm{d}^3 \kwv^\prime = \int \Theta^{(\text{type})}(\knorm, \knorm^\prime, \kangle) 2\pi \knorm^{\prime 2} \mathrm{d} \knorm^\prime \sin \kangle \mathrm{d} \kangle
\end{equation}

where $\text{type}=\mathrm{1ph},\ \mathrm{2ph}$ indicates the type of scattering. We also perform the radius integration  over $\knorm^\prime$ analytically to integrate the delta functions in Eq.~\ref{eq:Theta_1ph_integral} and ~\ref{eq:Theta_2ph_integral}.
The angular integrations in Eq.~\ref{eq:rates_int} are performed numerically; details are provided in  Sec.~\ref{sec:details}.

The  computational flow of the semi-analytical model is as follows.
First, we generate a grid of $\knorm$, $\kangle$, $\kmnorm$ and  calculate the corresponding $\knorm^\prime$ from the energy conservation conditions for each subprocess.
Second, we calculate $\Theta^{\mathrm{(1ph)}}(\knorm, \knorm^\prime, \kangle)$ by Eq.~\ref{eq:Theta_1ph_integral} and $\Gamma_{\knorm}^{(\mathrm{1ph})}$ by Eq.~\ref{eq:rates_int}.
Then, we calculate $\tilde{I}_\text{1}^{(\alpha_1, \alpha_2)}(\knorm, \knorm^\prime, \kangle, \kmnorm)$ by Eq.~\ref{eq:I_tilde}.
Finally, we perform the self-consistent 2ph iterations through Eq.~\ref{eq:int_I_spherical}, \ref{eq:int_I_itf}, \ref{eq:Theta_2ph_integral} and \ref{eq:rates_int} until convergence, where Eq.~\ref{eq:int_I_itf} is calculated every 10 iterations.

From the perspective of computational cost, the semi-analytical model reduces the number of integration variables in the 2ph scattering rate calculation from 9 (integration over $\kwv$, $\kwv^\prime$, $\km$) to 5 ($\knorm$, $\kangle$, $\km$) due to the spherical symmetry, and avoids the recomputation of $I_\text{1}^{(\alpha_1, \alpha_2)}$ in the 2ph self-consistent iterations due to the separation of radius and angular integration in Eq.~\ref{eq:int_I_spherical}.  This reduction allows for the use of a  denser grid for the intermediate state integration and thereby  reduces the discretization error. Therefore, the total scattering rates can be calculated with negligible discretization error compared with the ab-initio calculation.
However, since the semi-analytical model is only for $\Gamma$-$\Gamma$ scattering, a discretized scattering matrix ($\Theta_{\kwv^\prime, \kwv}$) is still needed to compute the drift velocity and current PSD which are affected by $\Gamma$-L intervalley scattering.
Therefore, the discretization error in the final state integration cannot be avoided for the present calculations.
Nevertheless, the semi-analytical model still decreases the discretization error of the intermediate state integration and treats the full 2ph scattering term. The differences between the ab-initio calculation and the semi-analytical model for $\Gamma$ intravalley scattering are summarized in Table~\ref{tab:table}.

\begin{table}[htbp]
    \begin{ruledtabular}
    \begin{tabular}{ccc}
         & Ab-initio calculation  & Semi-analytical model \\ \hline
      \multirow{2}{*}{Final state integration} & \multirow{2}{*}{200$\times$200$\times$200} & 200$\times$200$\times$200 for observables \\ 
                  &                   & Effectively exact for scattering rates \\
      Intermediate state  integration &  200$\times$200$\times$200 & Effectively exact \\
      Processes & On-shell only & All processes included \\ 
      Two-phonon iterations & 3 & Iterate until convergence
      \\
      Interference term & Not included & Included
      \\
      Computational time (CPU hours) & 50000 & 40  \\
    \end{tabular}
    \end{ruledtabular}
    \caption{Comparison between the ab-initio calculation and the semi-analytical model for $\Gamma-\Gamma$ intravalley scattering.  The semi-analytical model improves upon the ab-initio model in all respects except the final state integration grid density for observables, for which the same grid is used.}
    \label{tab:table}
\end{table}

\section{Computational details} \label{sec:details}


\subsection{Ab-initio calculations}
The ab-initio calculation parameters are identical to those in our previous work \cite{cheng2022high}. In brief,  electronic structure and e-ph matrix elements are computed using Density Functional Theory (DFT) and Density Functional Perturbation Theory (DFPT) with \textsc{Quantum Espresso} \cite{giannozzi_qe_2009, giannozzi_qe_2017} with an  $8 \times 8 \times 8$ coarse grid, 72 Ryd plane wave energy cutoff, a relaxed lattice parameter of 5.556 $\textrm{\AA}$, and a non-degenerate carrier concentration of $10^{15}$ cm$^{-3}$.
Following our previous work \cite{cheng2022high}, we apply a band structure correction for both the $\Gamma$ valley and the L valley.
For the $\Gamma$ valley, we use a spherically symmetric Kane model band structure \cite{conwell_1968} with an experimental effective mass of 0.067$m_e$ and a non-parabolicity of 0.64 Ryd$^{-1}$ \cite{Lundstrom_2000}. For the L valley, we shift the energy by 50 meV to achieve the experimental $\Gamma$-L valley separation of 300 meV.
Wannier interpolation in \perturbo~\cite{perturbo_2021} is then applied to interpolate the e-ph matrix elements to a finer grid of $200 \times 200 \times 200$.
After the e-ph matrix elements are obtained, the 1ph and 2ph scattering matrices are computed according to Eqs.~\ref{eq:Theta_1ph} and \ref{eq:Theta_2ph}.
The delta functions in Eqs.~\ref{eq:Theta_1ph} and \ref{eq:Theta_2ph_subprocess} are approximated by a Gaussian function  with a standard deviation of 5 meV.

During the computation of scattering matrices, a phonon frequency cutoff of 2 meV is applied to neglect phonons with low frequencies.
An energy cutoff of 360 meV above the conduction band minimum is used to reduce the number of k points in the Brillouin zone integration.
The 2ph calculation applies the on-shell approximation by restricting the off-shell extent to 25 meV.
Following Ref.~\cite{warmelectrons}, the linear systems of equations representing the Boltzmann equation are solved by the generalized minimal residual method (GMRES).

\subsection{Semi-analytical model}
The band structure used in the semi-analytical model is the same as that in the ab-initio calculation.
The  LO phonon energy is taken to be $\omega_\text{LO} = 35$ meV. The function $g_{\text{LO}}(\qnorm)$ is obtained by a weighted averaged of $g_\text{LO}(\kwv_i, \pwv_i)$:
\begin{equation}
g_{\text{LO}}(\qnorm)= \frac{\sum_i g_\text{LO}(\kwv_i, \pwv_i)\exp(-\frac{(\qnorm-\pnorm_i)^2}{2\sigma^2}{})}
{\sum_i \exp(-\frac{(\qnorm-\pnorm_i)^2}{2\sigma^2}{})}
\end{equation}
where the standard deviation $\sigma=5\times 10^{-4}$ Ryd, the summation is over all the on-shell processes, and the $g_\text{LO}(\kwv_i, \pwv_i)$ are calculated by the Wannier interpolation.
The spherical coordinates integration in Eq.~\ref{eq:I_tilde} is defined such that the the $\theta_{\kwv_m}=0$ direction is orthogonal with both $\kwv$ and $\kwv^\prime$.
200 grid points are used for each $\theta_{\kwv_m}$ and $\phi_{\kwv_m}$ integration.
The radius integration in Eq.~\ref{eq:int_I_spherical} is transformed into the integration of $\epsilon_{\kmnorm}$ and performed using an adaptive integration range with 120 grid points.
The integration range is $(\epsilon_{\knorm^\prime} + \alpha_2 \hbar \omega_0 - 6 \hbar \Gamma_{\kmnorm}, \epsilon_{\knorm^\prime} + \alpha_2 \hbar \omega_0 + 6 \hbar \Gamma_{\kmnorm})$, corresponding to a width of $12 \hbar \Gamma_{\kmnorm}$ around the center of the Lorentzian function in the denominator of Eq.~\ref{eq:int_I_spherical}. 
The relative residual error from this choice of integration limits is estimated as $1/12^2 \approx 0.7\%$.
In the final state integration of both 1ph and 2ph, the angular integrations in Eq.~\ref{eq:rates_int} are performed with  200 grid points. All the above numerical integrations are performed on uniform grids using the midpoint rule.
The 2ph calculation is performed with 20 iterations, with the interference term Eq.~\ref{eq:int_I_itf} updated every 10 iterations.
The relative difference between the 10th and 20th iterations is less than 1\%, indicating convergence of the iterative process. The discretization of $\Theta(\knorm, \knorm^\prime, \kangle)$ to $\Theta_{\kwv^\prime,\kwv}$ is performed by the regular grid interpolation provided in scipy \cite{2020SciPy-NMeth}.

\section{Results}

\subsection{ $\mathbf{\Gamma}$ valley scattering rates at 300 K} \label{sec:rates}

\begin{figure}
    \centering
    {
    \phantomsubcaption\label{fig:rate_300K-a}
    \phantomsubcaption\label{fig:rate_300K-b}
    \includegraphics[width=1\textwidth]{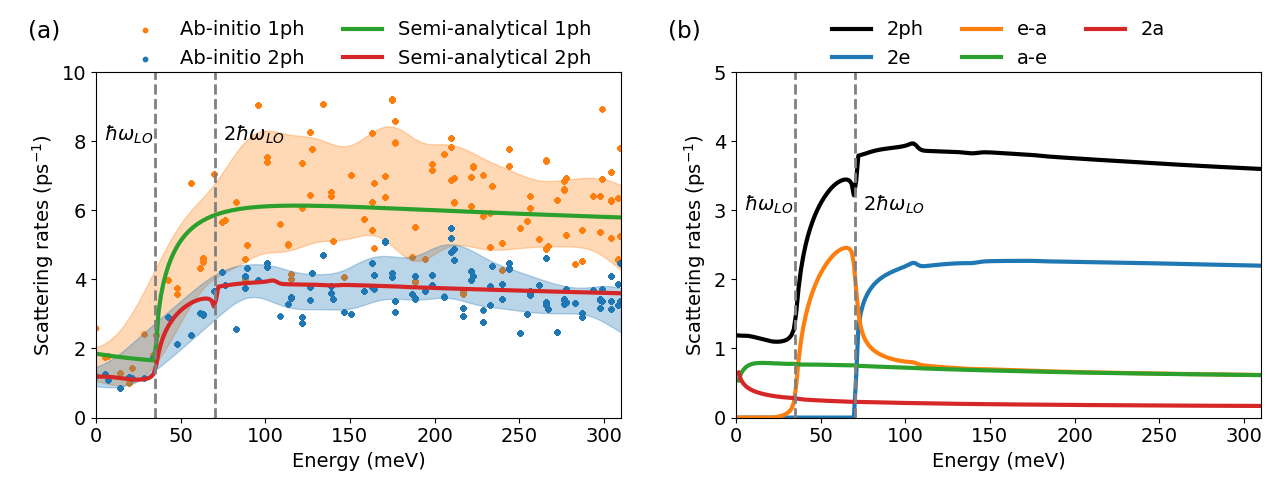}
    }
    \caption{ (a) 1ph and 2ph $\Gamma$-valley  scattering rates at 300 K obtained from ab-initio calculations (symbols and shaded regions) and semi-analytical model (lines).  Note that intervalley scattering is excluded from these rates. Due to the variations of the ab-initio rates, we apply a Gaussian smearing and plot the shaded region to indicate the region within a standard deviation. (b) Decomposition of  semi-analytical 2ph rates (black solid line) to four contributing subprocesses (blue, orange, green, and red solid lines). In both (a) and (b), vertical dashed lines indicating energies of $\hbar \omega_\text{LO}$ and $2\hbar \omega_\text{LO}$ are plotted.}
    \label{fig:rate_300K}
\end{figure}

We first present the 1ph and 2ph $\Gamma$ valley scattering rates versus energy  obtained by ab-initio calculation and the semi-empirical model for GaAs at 300K in Fig.~\ref{fig:rate_300K-a}. For both 1ph and 2ph rates, the  ab-initio calculations and the semi-analytical model are in quantitative agreement. Specifically, we observe a rapid increase of the 1ph and 2ph rates at $\hbar \omega_\text{LO} \approx 35$ meV followed by a nearly constant trend. The degree of agreement between  the semi-analytical model and the mean values of the ab-initio calculation is notable considering the semi-analytical model includes off-shell processes and the interference term, both of which are neglected in the ab-initio calculations. The agreement can be attributed in part to the cancellation of errors between the limitation on the iteration number and the on-shell approximation in the ab-initio calculation. The third iteration of the 2ph rates yields values that are overestimated from the converged value by about 9\%, while the on-shell approximation and the non-interference approximation are found to underestimate the rates by around 3\% and 5\%, respectively. These approximations offset each other to yield good agreement between the two approaches. Overall, the contribution of off-shell processes is found to make only a minor contribution to the $\Gamma$ intravalley 2ph scattering rates even up to energies of 300 meV.

The major difference between the ab-initio and the semi-analytical rates is the variation of the individual rates in the ab-initio calculation in a given energy range, which is due to the relatively low grid density used in the ab-initio calculation.
As explained in Sec.~\ref{sec:semi-analytical_model}, the semi-analytical model uses a significantly finer grid, leading to negligible variations in individual scattering rates in the same energy range.
Although anisotropy could in principle lead to similar variations of the ab-initio rates, this contribution is negligible owing to the high spherical symmetry of the band structure and e-ph matrix elements  (about 3\% as mentioned in Sec.~\ref{sec:semi-analytical_model}).

The high grid density in the semi-analytical model enables features in the scattering rates to be observed that cannot be discerned in the ab-initio calculations, including the previously mentioned rapid increase of 1ph and 2ph rates at $\hbar \omega_\text{LO}$, and also a small but evident kink at $2\hbar \omega_\text{LO}$ (about 70 meV).

We now analyze each of these observations. For the 1ph rates, the increase at $\hbar \omega_\text{LO}$ is because LO phonon emission from an electron may only occur above an energy of  $\hbar \omega_\text{LO}$.
For the 2ph rates, the situation is more complicated due to the existence of four subprocesses (2e, e-a, a-e, 2a) in 2ph scattering.
To better understand the features in the 2ph scattering rates, the scattering rates of the four subprocesses are plotted separately in Fig.~\ref{fig:rate_300K-b}.
We observe that the increase of the total 2ph rates at $\hbar \omega_\text{LO}$ can be attributed to the e-a subprocess due to a similar reason with the emission subprocess in 1ph process, namely that the e-a subprocess requires the electron to have energy exceeding $\hbar \omega_\text{LO}$.
The kink at $2\hbar \omega_\text{LO}$ comes from the cancellation between the increase of the 2e rates and the decrease of e-a rates. 
The increase of the 2e rates is due to the emission of 2 LO phonons at energies higher than $2\hbar \omega_\text{LO}$.
The decrease of the e-a rates is due to the increase of the intermediate state rates in the denominator of Eq.~\ref{eq:W_tilde} (or Eq.~\ref{eq:int_I_spherical}).
Specifically, for an e-a subprocess with $\epsilon_{\kwv}=2\hbar \omega_\text{LO}$, the corresponding intermediate state has the band energy $\epsilon_{\kwv+\qwv}=\hbar \omega_\text{LO}$, where an increase of both $\Gamma^{(\mathrm{1ph})}$ and $\Gamma^{(\mathrm{2ph})}$ occurs as explained above.
In fact, the effect of the intermediate state self-energy leads to change at any integer multiple of $\hbar \omega_\text{LO}$, but the magnitudes are decaying with increasing energy such that they cannot be observed at high energy.

Another observation in Fig.~\ref{fig:rate_300K-b} is the difference between the e-a rates and a-e rates asymptotically decreases to zero with increasing energy.
This can be understood by analysis of Eq.~~\ref{eq:Theta_2ph_integral}.
Specifically, in the process $\kwv_i \rightarrow \kwv_m \rightarrow \kwv_f$, the relative differences between norms of the state vectors $\knorm_i$, $\knorm_m$, $\knorm_f$ become small at  high energies (above a few multiples the LO phonon energy), so that the factor $I^{(\alpha_1, \alpha_2)}(\knorm, \knorm^\prime, \kangle)$ becomes insensitive to the subprocess type.
Thus, their magnitudes are fully determined by the phonon occupation factor $A_{\alpha_1}A_{\alpha_2}$ defined in Eq.~\ref{eq:phonon_occupation} which satisfy $\Gamma^{(2a)} / A_{-1}^2 = \Gamma^{(e-a)} / A_{-1}A_{+1} = \Gamma^{(a-e)} / A_{-1}A_{+1} = \Gamma^{(2e)} / A_{+1}^2$.
From this relationship, we find $\Gamma^{(e-a)} = \Gamma^{(a-e)}$.
Furthermore, a common ratio $\Gamma^{(2a)} / \Gamma^{(e-a)} = \Gamma^{(e-a)} / \Gamma^{(2e)} = A_{-1}/A_{+1}$ can also be obtained for the subprocess rates at the high energy region.
This relationship is observed in Fig.~\ref{fig:rate_300K-b}.

\subsection{Drift velocity and current PSD at 300 K}

\begin{figure}
    \centering
    {
    \phantomsubcaption\label{fig:drift_psd-a}
    \phantomsubcaption\label{fig:drift_psd-b}
    \includegraphics[width=1\textwidth]{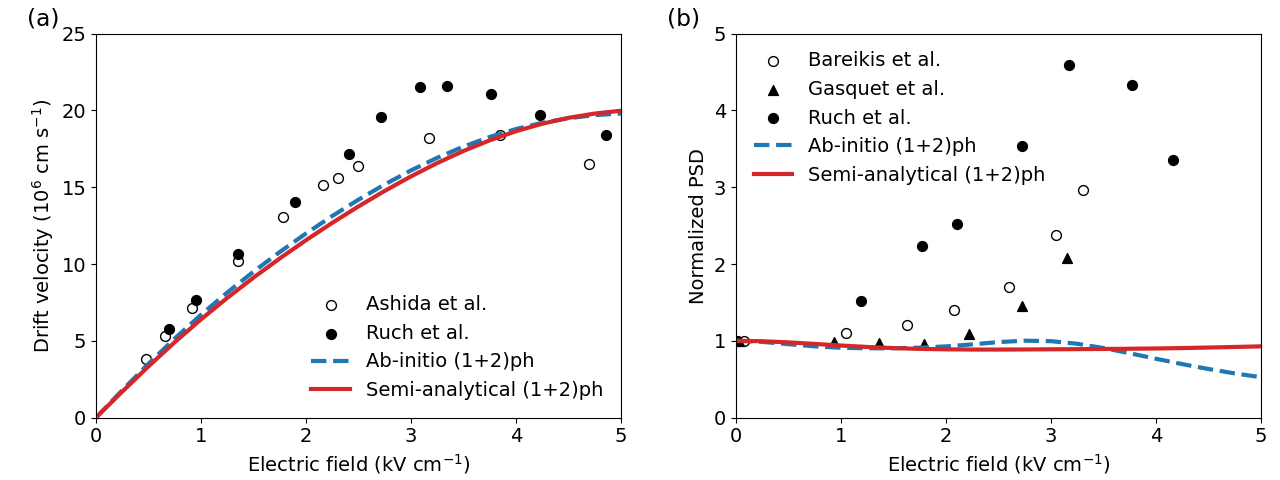}
    }
    \caption{Drift velocity and normalized PSD versus electric field for the (1+2)ph results obtained by ab-initio calculation (dashed blue line) and semi-analytical model (solid red line) at a temperature of 300 K. (a) Drift velocity versus electric field: The ab-initio calculation and semi-analytical model qualitatively agree with the measurements of Ruch et al. \cite{ruchkino1967} (filled circles) and Ashida et al. \cite{ashida1974} (open circles)  (b) Normalized PSD versus electric field. Values calculated using two approaches show consistent discrepancies compared with the results obtained from noise temperature and differential mobility measurements (filled circles, Ref.~\cite{gasquet1985} and open circles,  Ref.~\cite{Bareikis_1980}), and from time of flight experiments (triangles, Ref.~\cite{ruchkino1968}).}
    \label{fig:drift_psd}
\end{figure}

We now examine the transport and noise properties from each model.  For the semi-analytical model results, the $\Gamma\text{-}\Gamma$ block of the scattering matrix is calculated by the semi-analytical model in Eq.~\ref{eq:Theta_1ph_integral} and \ref{eq:Theta_2ph_integral}, while the $\Gamma$-L and L-L blocks are those of the ab-initio calculation. Figures \ref{fig:drift_psd-a} and \ref{fig:drift_psd-b} display the drift velocity and normalized current PSD, respectively, from the ab-initio calculations and the semi-analytical model. The experimental measurements are also plotted for comparison.
In Fig.~\ref{fig:drift_psd-a}, the ab-initio calculation and the semi-analytical model give a similar prediction of the drift velocity versus electric field up to 5 \eunits.
Both give the low-field mobility of around 7000 cm$^2$V$^{-1}$s$^{-1}$, which agrees with the experimental value at about 8000 cm$^2$V$^{-1}$s$^{-1}$ to within around 15\%. The similarity between the ab-initio and semi-analytical results is expected due to the agreement of their scattering rates as discussed in Sec.~\ref{sec:rates}.

In Fig.~\ref{fig:drift_psd-b}, the PSD obtained from different experimental measurements reveal a non-monotonic pattern characterized by an initial decrease, followed by a marked rise around the commencement of negative differential mobility, and a subsequent decrease. The origin of this trend was explained in Ref.~\cite{cheng2022high}.
However, both the ab-initio calculation and the semi-analytical model predict the PSD to be nearly independent of electric field and thus fail to predict the characteristic PSD peak at about 3 \eunits.

\subsection{Comparison of cryogenic $\mathbf{\Gamma}$ and $\mathbf{\Gamma}$-L scattering rates to experiment} \label{sec:helium}


\begin{figure}
    \centering
    {
    \phantomsubcaption\label{fig:helium-a}
    \phantomsubcaption\label{fig:helium-b}
    \includegraphics[width=1\textwidth]{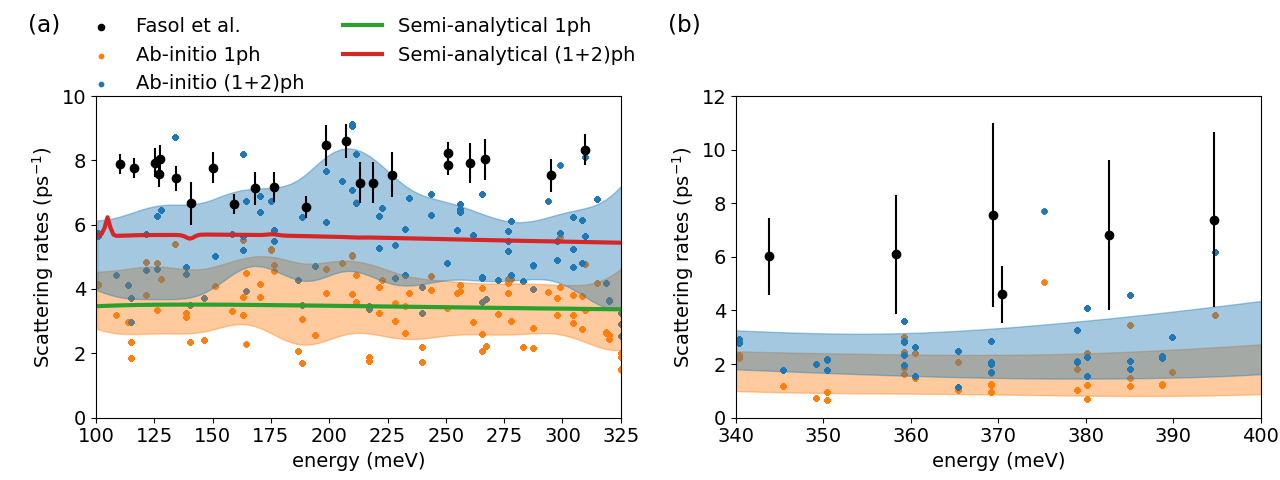}
    }
    \caption{Experimental (black symbols, Ref. \onlinecite{fasol1990}), ab-initio 1ph (orange symbols) and (1+2)ph (blue symbols), semi-analytical 1ph (green lines) and (1+2)ph results (red lines) of (a) $\Gamma$ valley scattering rates and (b) $\Gamma$-L intervalley scattering rates at helium temperatures. The (1+2)ph calculations agree better with experiment compared to the 1ph calculations in all cases. The (1+2)ph scattering rates agree with experiment to within 25\% and a factor of two for $\Gamma$ rates and $\Gamma$-L intervalley rates, respectively.}
    \label{fig:helium}
\end{figure}



The lifetimes of photoexcited hot electrons in GaAs have been experimentally measured at 10 K from an analysis of the linewidths of peaks from continuous-wave luminescence spectroscopy. \cite{fasol1990} In this section, we compare the ab-initio, semi-analytical and experimental scattering rates at cryogenic temperatures. In Fig.~\ref{fig:helium-a} and \ref{fig:helium-b}, we show the $\Gamma$ and $\Gamma$-L scattering rates obtained by ab-initio calculation, the semi-analytical model, and experiment. The calculations were performed at cryogenic temperatures to enable comparison with experiment. The experimental scattering rates and error bars are converted from the corresponding lifetimes and error bars directly reported in Ref.~\cite{fasol1990}.
Since the semi-analytical model is only valid for $\Gamma\text{-}\Gamma$ scattering, only the experimental and ab-initio results are shown in Fig.~\ref{fig:helium-b}.

Figure \ref{fig:helium-a} shows that the experimental and theoretical values for the $\Gamma$-valley scattering all yield a nearly constant value  between 100 meV and 325 meV. The experimental rates are about 8 ps$^{-1}$ in this energy range, while the semi-analytical 1ph and (1+2)ph calculations give about 3.5 ps$^{-1}$ and 6 ps$^{-1}$, respectively. This result affirms that the 2ph scattering makes a  non-negligible contribution to electron scattering in GaAs. Similarly with Sec.~\ref{sec:rates}, the ab-initio calculations give  mean values of the scattering rates that agree quantitatively with the semi-analytical model but with substantial scatter about the mean. Such observation again suggests that the approximations in the ab-initio calculations do not result in qualitative deviations.

Figure \ref{fig:helium-b} shows the $\Gamma$-L intervalley scattering rates of the ab-initio calculations and the photoluminescence experiments.
According to Ref.~\onlinecite{fasol1990}, the experimental $\Gamma$-L rates are obtained by 
\begin{equation} \label{eq:rates_GL}
    \Gamma_{\Gamma-L}=\Gamma_{\text{tot}}-\Gamma_{\Gamma-\Gamma},
\end{equation}
where $\Gamma_{\Gamma-\Gamma}$ is taken as a constant estimated by fitting the data in Fig.~\ref{fig:helium-a}.
To make a consistent comparison with the experimental estimation, the ab-initio $\Gamma$-L intervalley scattering rates in Fig.~\ref{fig:helium-b} are also calculated by Eq.~\ref{eq:rates_GL} instead of being directly calculated from the scattering matrix.
It is found that the experimental intervalley rates is about 6 ps$^{-1}$ in the energy range from 340 meV and 400 meV, while the ab-initio 1ph and (1+2)ph give around 2 ps$^{-1}$ and 3 ps$^{-1}$, respectively.
Although the additional 2ph calculation decreases the deviation from experiment results, an underestimation of a factor of two is still observed. This discrepancy could be attributed to the need for an off-shell extent larger than 25 meV in the 2ph calculation, owing to the larger intermediate state scattering rates (appearing in the denominator of Eq.~\ref{eq:W_tilde} and Eq.~\ref{eq:int_I_spherical}) at energies above 300 meV. However, at present a larger off-shell extent is computationally infeasible.

\section{Discussion}

The semi-analytical model treats the full scattering term for $\Gamma$ intravalley 2ph scattering but does not qualitatively alter transport and noise properties. In particular, the marked discrepancy with the experimental PSD remains. We now examine alternate possible origins for the discrepancy.

\subsection{Underestimated $\mathbf{\Gamma}$-L intervalley scattering rates}

A comparison of our computed cryogenic rates with those measured from photoluminescence experiments indicates that the  $\Gamma$ rates agree to within 25\%, but the $\Gamma$-L intervalley rates are underestimated by around a factor of two.  Despite this underestimate, prior work suggests that this effect is unlikely to reconcile the PSD discrepancy. Specifically, Monte Carlo simulations of electron transport in GaAs with a three-valley $\Gamma$-L-X model have found that increased intervalley scattering suppresses the PSD feature (see Fig. 7 of Ref.~\cite{povzela1980electron}). Therefore, although the possibility cannot be definitively excluded at present, including intervalley scattering processes beyond those treated already is not expected to yield improved agreement with the PSD.

\subsection{Contribution from simultaneous electron-two-phonon interaction} \label{sec:e-2ph}

According to Ref.~\onlinecite{cheng2022high}, another possibility is that the contribution of electron-two-phonon (e-2ph) interaction \cite{kocevar_1980} is not considered.
Here, we make a qualitative estimation of the magnitude of this effect based on the Fröhlich mechanism for electron scattering.
According to Ref.~\onlinecite{kocevar_1980}, the e-ph Hamiltonian up to the second order can be written as 
\begin{equation} \label{eq:eph-Hamiltonian}
    H_{\text{e-ph}} = \sum_{\boldsymbol{R}\kappa} \boldsymbol{u}_{\boldsymbol{R}\kappa} \cdot \nabla V(\boldsymbol{r}-\boldsymbol{R}_\kappa) + \frac{1}{2} \sum_{\kappa} \boldsymbol{u}_{\boldsymbol{R}\kappa} \cdot \nabla \nabla V(\boldsymbol{r}-\boldsymbol{R}_\kappa) \cdot \boldsymbol{u}_{\boldsymbol{R}\kappa},
\end{equation}
where $\kappa$ is the index of atom in a unit cell, $\boldsymbol{R}$ is the unit cell position, $\boldsymbol{u}_{\boldsymbol{R}\kappa}$ is the corresponding phonon-induced displacement, and $V$ is the electron potential. In the long wavelength limit, the electric potential for the Fr{\"o}hlich interaction can be obtained by assigning a point dipole to each atom \cite{verdi2015frohlich}. For acoustic phonons with the same displacements for atoms in the same unit cell, the net dipole moment will be zero and no scattering will occur. 
Similarly, the electric potential for e-2ph interaction can be obtained by assigning a point quadrupole to each atom.
Following the same logic, the net quadrupole moment will be zero  if the quadrupoles are induced by two acoustic phonons or two optical phonons, which means that the simultaneous e-2ph interaction based on the Fr{\"o}hlich interaction can only be induced by an acoustic and optical phonon.


We estimate the order of magnitude of such simultaneous e-2ph interaction involving an acoustic and optical phonon. A full derivation can be found in the Appendix.
The final estimated scattering rates in the $\Gamma$ valley is
\begin{equation} \label{eq:e-2ph_rates}
\begin{aligned}
    \Gamma^{(\mathrm{e-2ph})}(\knorm) &\sim 
    \frac{8\pi^2}{\Omega_\text{BZ} \hbar}  \frac{\knorm^3}{\epsilon_{\knorm}} \left(\frac{eZ }{\Omega\epsilon_\infty} \sqrt{\frac{\hbar}{2M\omega_\text{A}}} \sqrt{\frac{\hbar}{2M\omega_\text{O}}}\right)^2 (N_\text{A} + 1) (N_\text{O} + 1),
\end{aligned}
\end{equation}
where $\omega_\text{O}$ and $\omega_\text{A}$ are frequencies of optical and acoustic phonons at the edge of Brillouin zone, $N_\text{O}$ and $N_\text{A}$ are the corresponding phonon occupations, $\Omega_\text{BZ}$ is the Brillouin zone volume, $M$ is the average atom mass in a unit cell, $\epsilon_\infty$ is the high-frequency permittivity, and $Z$ is the Born effective charge of a single atom.
For a typical $\knorm$ such that $\epsilon_{\knorm}=200$ meV, the e-2ph scattering rates at 300 K can be estimated as $10^{-2.5}$ ps\textsuperscript{-1}, which is about 3.5 orders of magnitude smaller than the ab-initio or semi-analytical 2ph scattering rates obtained in this work. Thus, we conclude that the effect of the simultaneous e-2ph interaction based on the Fr\"{o}hlich interaction can be neglected.


\subsection{Space charge domains and experimental non-idealities}

Finally, we consider an alternate mechanism for the PSD peak which does not rely solely on intervalley scattering. The earliest studies of negative differential resistance in GaAs arose from the observation of current instabilities at electric fields approaching a threshold value of around  3 \eunits. \cite{gunn1964} These instabilities were  attributed to the formation of space charge domains associated with the negative differential resistance. The typical Boltzmann formalism used to describe charge transport from first principles would not include the contribution of such effects because it neglects real-space gradients and space charge effects which are essential to the instability.

Space charge instabilities manifest as current fluctuations, and therefore the nucleation of space-charge domains could explain the PSD peak around the threshold field. However, an inconsistency with this explanation is that the increase in PSD begins at a field below the threshold value for NDR as in Fig~\ref{fig:drift_psd}. This inconsistency can be accounted for by considering the possibility that the local electric field exceeds the threshold even though the average field does not. Such a possibility was investigated theoretically by McCumber and Chynoweth \cite{mccumber_1966}, who found that the dipole layer generation process was sensitive to inhomogeneities such as doping fluctuations that would arise from purely random Poisson statistics. The dipole layer was found to propagate  even if the average uniform field was less than the nominal threshold field based on the velocity-field characteristics.

These considerations support the explanation of the PSD peak in terms of  instabilities associated with the local electric field exceeding the threshold field for negative differential resistance space charge domain formation. This finding has implications for the use of transport studies to determine intervalley scattering strength. In particular, Monte Carlo methods have been used for decades for this purpose in GaAs by interpreting transport and noise measurements, and in those simulations, noise was assumed to arise from solely intervalley scattering. Our result indicates that this approach to studying intervalley scattering is not valid  because the physical origin of noise differs from that assumed in the model. Instead, photoluminescence methods which directly provide an electronic lifetime as in Ref.~\cite{fasol1990} should be employed.


\section{Conclusions}
We have introduced a semi-analytical model of 1ph and 2ph $\Gamma$ intravalley scattering for electrons in GaAs which allows for prior approximations in the treatment of the 2ph term for $\Gamma$ scattering to be lifted while incurring errors of  a few percent. We find that the calculated transport and noise properties are qualitatively unchanged from the ab-initio values. The computed drift velocity agrees with experiment to within 15\%, while agreement with measured cryogenic scattering rates are within 25\% for the $\Gamma$ valley scattering rates and a factor of two for  $\Gamma$-L intervalley scattering. However, the qualitative discrepancy for the  PSD  is not improved with the semi-analytical model. Considering these observations and prior work, we conclude that the PSD peak  arises from space charge domain formation rather than partition noise associated with intervalley scattering, as has been assumed for many decades. This result implies that care must be taken when  interpreting transport and electrical noise measurements in terms of intervalley scattering. Our findings highlight the insights into charge transport that can be obtained from a first-principles treatment of high-field charge transport and noise properties.



\section{Acknowledgements}

This work was supported by AFOSR under Grant Number FA9550-19-1-0321.
The authors thank P.~Cheng, B.~Hatanp\"a\"a, D.~Catherall, S-N.~Sun and T.~Esho for helpful discussions.

\bibliography{references}

\clearpage

\appendix*

\section{Derivation of simultaneous electron-two-phonon scattering rates} \label{sec:appendix}

We provide a  derivation for estimation of the simultaneous e-2ph scattering rates given in Sec.~\ref{sec:e-2ph}. Consider a crystal in which each primitive unit cell has atoms with charge $Z_{\kappa}$ at position $\boldsymbol{\tau}_{\kappa}$, \textbf{$\boldsymbol{R}$} is the lattice vector, and $\boldsymbol{G}$ is the reciprocal lattice constant. The lattice displacement $u_{\boldsymbol{R}\kappa}$ is decomposed using normal modes: 
\begin{equation} \label{eq:phonon_decomposition}
u_{\boldsymbol{R}\kappa}=\sum_{\boldsymbol{q}\nu}\boldsymbol{u}_{\boldsymbol{q}\kappa\nu}e^{i\boldsymbol{q}\cdot\boldsymbol{R}},
\end{equation}
where 
\begin{equation}
\boldsymbol{u}_{\boldsymbol{q}\kappa\nu}=\sqrt{\frac{\hbar}{2NM_{\kappa}\omega_{\boldsymbol{q}\nu}}}\boldsymbol{e}_{\boldsymbol{q}\kappa\nu}(b_{\boldsymbol{q}\nu}^{\dagger}+b_{\boldsymbol{-q}\nu}),
\end{equation}
$N$ is the number of unit cells in a supercell, $M_\kappa$ is the mass of atom $\kappa$, $\nu$ is the phonon mode, 
$\boldsymbol{e}_{\boldsymbol{q}\kappa\nu}$ is the phonon polarization unit vector, $b_{\boldsymbol{q}\nu}^{\dagger}, and b_{\boldsymbol{q}\nu}$ are creation and annihilation operators of phonon $\boldsymbol{q}\nu$, respectively.

The Coulomb potential energy of an electron generated by point charge $Z$ at position $\boldsymbol{R}_{\kappa}=\boldsymbol{R}+\boldsymbol{\tau}_{\kappa}$ is
\begin{equation}
V(\boldsymbol{r}-\boldsymbol{R}-\boldsymbol{\tau}_{\kappa}) =-\frac{Z_{\kappa}e^{2}}{4\pi\epsilon_{\infty}|\boldsymbol{r}-\boldsymbol{R}-\boldsymbol{\tau}_{\kappa}|}.
\end{equation}
It will be convenient to rewrite $V(\boldsymbol{r}-\boldsymbol{R}-\boldsymbol{\tau}_{\kappa})$ in reciprocal space by Fourier transformation:
\begin{equation} \label{eq:fourier}
V(\boldsymbol{r}-\boldsymbol{R}-\boldsymbol{\tau}_{\kappa}) =-\frac{1}{N\Omega}Z_{\kappa}\sum_{\boldsymbol{q}}\sum_{\boldsymbol{G}}V(\boldsymbol{q}+\boldsymbol{G)}e^{i(\boldsymbol{q}+\boldsymbol{G})\cdot(\boldsymbol{r}-\boldsymbol{R}-\boldsymbol{\tau}_{\kappa})},
\end{equation}
where $\Omega$ is the primitive unit  cell volume, and
\begin{equation}
V(\boldsymbol{q}+\boldsymbol{G)}=\frac{e^{2}}{\epsilon_{\infty}(\boldsymbol{q}+\boldsymbol{G})^{2}}.
\end{equation}
Following Ref.~\onlinecite{kocevar_1980}, the electron-two-phonon Hamiltonian is
\begin{equation}
H(\boldsymbol{r})=\frac{1}{2}\sum_{\boldsymbol{R}\kappa}\boldsymbol{u}_{\boldsymbol{R}\kappa}\cdot\nabla\nabla V(\boldsymbol{r}-\boldsymbol{R}-\boldsymbol{\tau}_{\kappa})\cdot\boldsymbol{u}_{\boldsymbol{R}\kappa}.
\end{equation}
From Eq.~\ref{eq:fourier}, we can obtain
\begin{equation} \label{eq:2nd_derivative}
\nabla\nabla V(\boldsymbol{r}-\boldsymbol{R}-\boldsymbol{\tau}_{\kappa})=Z_{\kappa}(\boldsymbol{k}+\boldsymbol{G})(\boldsymbol{k}+\boldsymbol{G})\sum_{\boldsymbol{k}}\sum_{\boldsymbol{G}}V(\boldsymbol{k}+\boldsymbol{G)}e^{i(\boldsymbol{k}+\boldsymbol{G})\cdot(\boldsymbol{r}-\boldsymbol{R}-\boldsymbol{\tau}_{\kappa})},
\end{equation}
where the product of the vectors in this and following equations is defined as the outer product. By using Eq.~\ref{eq:phonon_decomposition} and Eq.~\ref{eq:2nd_derivative}, we have
\begin{equation} \label{eq:final_H}
\begin{aligned}
H(\boldsymbol{r}) & =\frac{1}{2\Omega}\sum_{\boldsymbol{R}\kappa}Z_{\kappa}\boldsymbol{u}_{\boldsymbol{R}\kappa}\cdot\nabla\nabla V(\boldsymbol{r}-\boldsymbol{R}-\boldsymbol{\tau}_{\kappa})\cdot\boldsymbol{u}_{\boldsymbol{R}\kappa}\\
 & =\frac{1}{N\Omega}\sum_{\boldsymbol{R}\kappa}\sum_{\boldsymbol{q}\nu<\boldsymbol{q}^{\prime}\nu^{\prime}}Z_{\kappa}\sum_{\boldsymbol{k}}\sum_{\boldsymbol{G}}\boldsymbol{u}_{\boldsymbol{q}\kappa\nu}\cdot(\boldsymbol{k}+\boldsymbol{G})\boldsymbol{u}_{\boldsymbol{q}^{\prime}\kappa\nu^{\prime}}\cdot(\boldsymbol{k}+\boldsymbol{G})V(\boldsymbol{k}+\boldsymbol{G)}e^{i(\boldsymbol{q}+\boldsymbol{q}^{\prime})\cdot\boldsymbol{R}}e^{i(\boldsymbol{k}+\boldsymbol{G})\cdot(\boldsymbol{r}-\boldsymbol{R}-\boldsymbol{\tau}_{\kappa})}\\
 & =\frac{1}{N\Omega}\sum_{\boldsymbol{G}\kappa}\sum_{\boldsymbol{q}\nu<\boldsymbol{q}^{\prime}\nu^{\prime}}Z_{\kappa}\boldsymbol{u}_{\boldsymbol{q}\kappa\nu}\cdot(\boldsymbol{q}+\boldsymbol{q}^{\prime}+\boldsymbol{G})\boldsymbol{u}_{\boldsymbol{q}^{\prime}\kappa\nu^{\prime}}\cdot(\boldsymbol{q}+\boldsymbol{q}^{\prime}+\boldsymbol{G})V(\boldsymbol{q}+\boldsymbol{q}^{\prime}+\boldsymbol{G})e^{i(\boldsymbol{q}+\boldsymbol{q}^{\prime}+\boldsymbol{G})\cdot(\boldsymbol{r}-\boldsymbol{\tau}_{\kappa})}\\
 & =\frac{e^{2}}{\Omega\epsilon_{\infty}}\sum_{\boldsymbol{G}\kappa}\sum_{\boldsymbol{q}\nu<\boldsymbol{q}^{\prime}\nu^{\prime}}Z_{\kappa}\boldsymbol{u}_{\boldsymbol{q}\kappa\nu}\cdot\frac{(\boldsymbol{q}+\boldsymbol{q}^{\prime}+\boldsymbol{G})(\boldsymbol{q}+\boldsymbol{q}^{\prime}+\boldsymbol{G})}{|(\boldsymbol{q}+\boldsymbol{q}^{\prime}+\boldsymbol{G})|^{2}} \cdot \boldsymbol{u}_{\boldsymbol{q}^{\prime}\kappa\nu^{\prime}} e^{i(\boldsymbol{q}+\boldsymbol{q}^{\prime}+\boldsymbol{G})\cdot(\boldsymbol{r}-\boldsymbol{\tau}_{\kappa})}\\
 & \sim\frac{e^{2}}{\Omega\epsilon_{\infty}}\sum_{\boldsymbol{G}\kappa}\sum_{\boldsymbol{q}\nu<\boldsymbol{q}^{\prime}\nu^{\prime}}Z_{\kappa}u_{\boldsymbol{q}\kappa\nu}u_{\boldsymbol{q}^{\prime}\kappa\nu^{\prime}}e^{i(\boldsymbol{q}+\boldsymbol{q}^{\prime}+\boldsymbol{G})\cdot(\boldsymbol{r}-\boldsymbol{\tau}_{\kappa})}(b_{\boldsymbol{q}\nu}^{\dagger}+b_{\boldsymbol{-q}\nu})(b_{\boldsymbol{q}^{\prime}\nu^{\prime}}^{\dagger}+b_{\boldsymbol{-q}^{\prime}\nu^{\prime}})\\
 & \sim\frac{e^{2}}{\Omega\epsilon_{\infty}}\sum_{\boldsymbol{q}\nu<\boldsymbol{q}^{\prime}\nu^{\prime}}\sum_{\kappa}Z_{\kappa}u_{\boldsymbol{q}\kappa\nu}u_{\boldsymbol{q}^{\prime}\kappa\nu^{\prime}}e^{i(\boldsymbol{q}+\boldsymbol{q}^{\prime})\cdot(\boldsymbol{r}-\boldsymbol{\tau}_{\kappa})}(b_{\boldsymbol{q}\nu}^{\dagger}+b_{\boldsymbol{-q}\nu})(b_{\boldsymbol{q}^{\prime}\nu^{\prime}}^{\dagger}+b_{\boldsymbol{-q}^{\prime}\nu^{\prime}})
\end{aligned}
\end{equation}
where $u_{\boldsymbol{q}\kappa\nu}=\sqrt{\hbar/2M_{\kappa}\omega_{q}}$ is the amplitude of the phonon displacement,
\begin{equation}
\sum_{\boldsymbol{R}}e^{i(\boldsymbol{q}+\boldsymbol{q}^{\prime}-\boldsymbol{k}-\boldsymbol{G})\cdot\boldsymbol{R}} =N\delta_{\boldsymbol{k},\boldsymbol{q}+\boldsymbol{q}^{\prime}}
\end{equation}
was used in the third line, 
\begin{equation}
\frac{(\boldsymbol{q}+\boldsymbol{q}^{\prime}+\boldsymbol{G})(\boldsymbol{q}+\boldsymbol{q}^{\prime}+\boldsymbol{G})}{|(\boldsymbol{q}+\boldsymbol{q}^{\prime}+\boldsymbol{G})|^{2}}\sim1
\end{equation}
was used in the fourth line, and the contribution of $\boldsymbol{G}\neq0$ are neglected from $\sum_{\boldsymbol{G}}$ in the last line.

We observe that the quadrupole moment $\sum_{\kappa} Z_{\kappa}\boldsymbol{u}_{\boldsymbol{q}\kappa\nu}\boldsymbol{u}_{\boldsymbol{q}^{\prime}\kappa\nu^{\prime}}$ appears in the third line of Eq.~\ref{eq:final_H}. 
For a polar material like GaAs that each unit cell has two atoms with opposite charge and similar mass, we can perform a $\mathbb{Z}_2$ symmetry analysis for the quadrupole moment.
Since $Z_{\kappa}$ has odd
symmetry for the two atoms in a unit cell, the phonon modes must be
one of odd symmetry (optical mode) and one of even symmetry (acoustic mode)
to avoid cancellation.

The electron-phonon scattering matrix elements can be written as
\begin{equation}
g_{\nu\nu^{\prime}}^{(2)}(\boldsymbol{q},\boldsymbol{p})\sim\frac{Z_{\kappa}e^{2}}{\Omega\epsilon_{\infty}}u_{\boldsymbol{q}\kappa\nu}u_{\boldsymbol{p}\kappa\nu^{\prime}},
\end{equation}
where the phase factor is neglected.

Similar to Eq.~\ref{eq:Theta_2ph_integral} and Eq.~\ref{eq:int_I}, we can derive the e-2ph collision matrix element for a specific phonon mode and subprocess type as
\begin{equation} \label{eq:Theta_def}
\begin{aligned}
\Theta_{\boldsymbol{k}^{\prime},\boldsymbol{k}} & \sim\frac{2\pi}{\hbar}\frac{1}{\Omega_{\text{BZ}}^{2}}\delta(\epsilon_{k^{\prime}}-\epsilon_{k}-\Delta E)\int_{\boldsymbol{q}+\boldsymbol{p}=\boldsymbol{k}^{\prime}-\boldsymbol{k}}N_{q}N_{p}\big|g^{(2)}(\boldsymbol{q},\boldsymbol{p})\big|^{2}d^{3}\boldsymbol{p}.
\end{aligned}
\end{equation}

In the following, we assume that phonon $\boldsymbol{q}$ is optical and phonon
$\boldsymbol{p}$ is acoustic. We additionally assume that optical phonons have no dispersion and acoustic phonons have linear dispersion with velocity $s$ such that:
\begin{equation}
\begin{aligned}
\omega_{q} & =\omega_{\text{O}}\\
\omega_{p} & \sim sp\sim\frac{p}{k_{\max}}\omega_{\text{A}},
\end{aligned}
\end{equation}
where $\omega_{\text{O}}$ and $\omega_{\text{A}}$ are phonon frequencies
at the edge of the Brillouin zone for optical and acoustic phonons, respectively. Since the frequencies of transverse and longitudinal phonons at the edge of the Brillouin zone are of the same magnitude here we neglect their difference. Using the phonon dispersion relation assumed above, we then have
\begin{equation}
\begin{aligned}
N_{q} & \sim N(\omega_{\text{O}})\\
N_{p} & \sim N(\omega_{\text{A}})\frac{N(\omega_{p})}{N(\omega_{A})}\sim N(\omega_{\text{A}})\frac{\omega_{A}}{\omega_{p}}\sim N(\omega_{\text{A}})\frac{k_{\max}}{p},
\end{aligned}
\end{equation}
where in the second line we assume the temperature is not too low in the sense of $\beta\hbar\omega<1$ or $\beta\hbar\omega\sim1$ so that $N(\omega)=\frac{1}{e^{\beta\hbar\omega}-1}\sim\frac{1}{\beta\hbar\omega}$,
and $k_{\max}$ is the wave vector the edge of the Brillouin zone. 

Similarly, we have
\begin{equation}
g^{(2)}(\boldsymbol{q},\boldsymbol{p})\sim g^{(2)}(k_{\max},k_{\max})\sqrt{\frac{k_{\max}}{p}},
\end{equation}
where 
\begin{equation}
g^{(2)}(k_{\max},k_{\max})=\frac{eZ}{\Omega\epsilon}\sqrt{\frac{\hbar}{2M\omega_{\text{A}}}}\sqrt{\frac{\hbar}{2M\omega_{\text{O}}}}.
\end{equation}

We can then calculate $\Theta_{\boldsymbol{k}^{\prime},\boldsymbol{k}}$
from Eq.~\ref{eq:Theta_def} as
\begin{equation}
\begin{aligned}
\Theta_{\boldsymbol{k}^{\prime},\boldsymbol{k}} & \sim\frac{2\pi}{\hbar}\frac{1}{\Omega_{\text{BZ}}^{2}}\big|g^{(2)}(k_{\max},k_{\max})\big|^{2}\delta(\epsilon_{k^{\prime}}-\epsilon_{k}-\Delta E)A_{\alpha_{1}}(\omega_{\text{O}})A_{\alpha_{2}}(\omega_{\text{A}})\int\Big(\frac{k_{\max}}{p}\Big)^{2}d^{3}\boldsymbol{p}\\
 & \sim\frac{2\pi}{\hbar}\frac{1}{\Omega_{\text{BZ}}^{2}}\big|g^{(2)}(k_{\max},k_{\max})\big|^{2}\delta(\epsilon_{k^{\prime}}-\epsilon_{k}-\Delta E)A_{\alpha_{1}}(\omega_{\text{O}})A_{\alpha_{2}}(\omega_{\text{A}})\Omega_{\text{BZ}}\\
 & = \frac{2\pi}{\hbar}\frac{1}{\Omega_{\text{BZ}}}\big|g^{(2)}(k_{\max},k_{\max})\big|^{2}\delta(\epsilon_{k^{\prime}}-\epsilon_{k}-\Delta E)A_{\alpha_{1}}(\omega_{\text{O}})A_{\alpha_{2}}(\omega_{\text{A}}),
 \end{aligned}
\end{equation}
where $\Omega_{\text{BZ}}$ is the Brillouin zone volume, $\alpha_{1,2}$ indicates whether a phonon is absorbed or emitted,
and
\begin{equation}
A_{\alpha}(\omega)=N(\omega)+\delta_{\alpha,+1}.
\end{equation}

The scattering rate can be calculated by integrating $\Theta_{\boldsymbol{k}^{\prime},\boldsymbol{k}}$ over $\boldsymbol{k}^{\prime}$:
\begin{equation}
\begin{aligned}
\Gamma_{k} & =\int\Theta_{\boldsymbol{k}^{\prime},\boldsymbol{k}}d^3 \boldsymbol{k}^{\prime}\\
 & \sim\int\frac{2\pi}{\hbar}\frac{1}{\Omega_{\text{BZ}}}\big|g^{(2)}(k_{\max},k_{\max})\big|^{2}\delta(\epsilon_{k^{\prime}}-\epsilon_{k}-\Delta E)A_{\alpha_{1}}(\omega_{\text{O}})A_{\alpha_{2}}(\omega_{\text{A}})4\pi k^{\prime^{2}}dk^{\prime}\\
 & =\frac{8\pi^{2}}{\hbar\Omega_{\text{BZ}}}\big|g^{(2)}(k_{\max},k_{\max})\big|^{2}A_{\alpha_{1}}(\omega_{\text{O}})A_{\alpha_{2}}(\omega_{\text{A}})k^{\prime^{2}}\frac{dk^{\prime}}{d\epsilon_{k^{\prime}}}\\
 & \sim\frac{8\pi^{2}}{\hbar\Omega_{\text{BZ}}}\big|g^{(2)}(k_{\max},k_{\max})\big|^{2}A_{\alpha_{1}}(\omega_{\text{O}})A_{\alpha_{2}}(\omega_{\text{A}})\frac{k^{3}}{\epsilon_{k}}\\
 & \sim\frac{8\pi^{2}}{\hbar}(\frac{k}{k_{\max}})^{3}\big|g^{(2)}(k_{\max},k_{\max})\big|^{2}A_{\alpha_{1}}(\omega_{\text{O}})A_{\alpha_{2}}(\omega_{\text{A}})\frac{1}{\epsilon_{k}}
\end{aligned}
\end{equation}
where $k^{\prime}\sim k$ is assumed since the phonon energy ($\leq35$ meV)
is low compared with the energy range we are considering ($\sim200$ meV).

Considering GaAs at temperature $\sim 300$ K and an energy  of about
$200$ meV (corresponding to $\frac{k}{k_{\max}} \sim 0.05$, we
have:
\begin{equation}
\begin{aligned}
A_{\alpha_{1}}(\omega_{\text{O}})\sim A_{\alpha_{2}}(\omega_{\text{A}})&\sim1
\\
\left(\frac{k}{k_{\max}}\right)^{3}&\sim0.05^{3}\sim10^{-4}
\\
\epsilon_{k}&\sim\text{0.2eV}\sim10^{-2}\text{Ry}
\\
\big|g^{(2)}(k_{\max},k_{\max})\big|^{2}&\sim10^{-4}\text{Ry}
\end{aligned}
\end{equation}

Thus
\begin{equation}
\Gamma_{k}\sim\frac{8\pi^{2}}{\hbar}\left(\frac{k}{k_{\max}}\right)^{3}\big|g^{(2)}(k_{\max},k_{\max})\big|^{2}A_{\alpha_{1}}(\omega_{\text{O}})A_{\alpha_{2}}(\omega_{\text{A}})\frac{1}{\epsilon_{k}}\sim10^{-8}\text{Ry}\sim10^{-4}\text{ps}^{-1}
\end{equation}

Considering $3\times3$ phonon polarizations and $2\times2$ subprocess types, the total scattering rate is about $\Gamma_{k}^{(\text{total})}\sim36\Gamma_{k}\sim10^{-2.5}\text{ps}^{-1}$, which is about 3.5 magnitudes lower than the 2ph rates studied in
this work.

\end{document}